\documentclass[prl,superscriptaddress,twocolumn,floatfix]{revtex4}
\usepackage{graphicx}
\usepackage{color}
\usepackage{dcolumn}
\usepackage{amsmath}
\usepackage{amssymb}

\newcommand{\lbco}{$\rm La_{2-{\it x}}Ba_{\it x}CuO_4$}
\newcommand{\lbcoe}{$\rm La_{1.875}Ba_{0.125}CuO_4$}

\newcommand{\sus}{susceptibility}

\begin{document}

\title{Spontaneous symmetry breaking by charge stripes in the high-pressure phase of superconducting La$_{1.875}$Ba$_{0.125}$CuO$_4$}
\author{M. H\"ucker}
\affiliation{Brookhaven National Laboratory, Upton, New York 11973-5000, USA}
\author{M. v. Zimmermann}
\affiliation{Hamburger Synchrotronstrahlungslabor HASYLAB at Deutsches Elektronen-Synchrotron, 22603 Hamburg, Germany}
\author{M. Debessai}
\author{J. S. Schilling}
\affiliation{Department of Physics, Washington University, St. Louis, Missouri 63130, USA}
\author{J. M. Tranquada}
\author{G. D. Gu}
\affiliation{Brookhaven National Laboratory, Upton, New York 11973-5000, USA}

\date{\today}

\begin{abstract}
In those cases where charge stripe order has been observed in cuprates, the crystal structure is such that the average rotational symmetry of the CuO$_2$ planes is reduced from four-fold to two-fold.  As a result, one could argue that the reduced lattice symmetry is essential to the existence of stripe order.  We use pressure to restore the average four-fold symmetry in a single crystal of \lbcoe, and show by x-ray diffraction that charge stripe order still occurs.  Thus, electronically-driven stripe order can spontaneously break the lattice symmetry.
\end{abstract}

\maketitle

Charge and spin stripe order has been observed in a limited class of cuprate superconductors \cite{tran95a,fuji04,fink09}.  An enduring question is whether stripe correlations represent a fundamental instability of hole-doped CuO$_2$ planes \cite{zaan01,kive03}, which could be relevant to the unconventional superconductivity, or whether stripes are the consequence of a particular lattice structure with only two-fold symmetry of the planes, in which case they would represent a less interesting state that simply competes with bulk superconductivity.  There have been theoretical proposals for dynamic electronic correlations that should intrinsically break the four-fold symmetry of the planes \cite{kive98,vojt08,huh08}. Intriguing observations of anisotropic spin \cite{hink08} and transport \cite{ando02,daou09} properties in underdoped YBa$_2$Cu$_3$O$_{6+x}$ have been reported; however, the structural symmetry reduction due to Cu-O chains has motivated alternative explanations \cite{yama09,sush09}.  The observation of spontaneous symmetry breaking by stripe order in an otherwise square lattice would resolve the significance of stripes.  In the present work, we use high pressure to tune the crystal structure of \lbcoe, restoring  four-fold symmetry to the planes, and apply x-ray diffraction to demonstrate that, indeed, charge-stripe order still develops.  Our results provide strong evidence that stripe correlations in the cuprates are electronically driven and do not depend on reduced lattice symmetry.

At ambient pressure and room temperature, \lbcoe\ has the high-temperature tetragonal (HTT) structure, with 4-fold symmetric planes.  On cooling below $T_{\rm HT}$, the structure transforms to the low-temperature orthorhombic (LTO) phase, and below $T_{\rm LT}$ one reaches the low-temperature tetragonal (LTT) phase \cite{axe89}.  In each of the latter two phases, the CuO$_2$ planes have only 2-fold symmetry due to tilts of the Cu-O octahedra about an in-plane axis.  In the LTT phase, the tilt axis is along a Cu-O bond direction, so that orthogonal Cu-O in-plane bonds are inequivalent.  Previous diffraction studies have shown that $T_{\rm HT}$ and $T_{\rm LT}$ decrease with pressure \cite{craw05,kata93}; $T_{\rm HT}$ reaches zero at a critical pressure, $p_c$.  

The in-plane anisotropy of the LTT phase pins charge stripes \cite{tran95a,vonz98,fuji04}.  If the crystallographic anisotropy drives the charge order, then we would expect the charge order (CO) to disappear at $p_c$.   Though experimentally challenging, we can test this possibility by directly monitoring the charge order with {\it in situ} x-ray diffraction.  Another quantity that is sensitive to stripe order is the bulk superconducting transition temperature, $T_c$, which is strongly depressed when stripe order is optimal \cite{tran97a}.   If bond-aligned charge-stripe order can only occur in the LTT phase, then one would expect to see a large jump in $T_c$ when the LTT phase is suppressed.  Previous studies have indicated a modest, continuous increase in $T_c$ on suppressing the LTT phase, though $T_c$ remains lower than one might anticipate \cite{kata93,craw05,yama92b}.  We are not aware of any previous attempts to directly measure charge stripe order under pressure, although there have been recent high-pressure studies of charge-density-wave order in Cr \cite{jara09} and in $R$Te$_3$ ($R=$La, Ce) \cite{sacc09}.

The \lbcoe\ single crystal studied here and those measured in earlier work \cite{li07,huck08} were cut from the same large crystal. A mosaic spread of 0.008$^\circ$ has been determined at the (110) Bragg reflection in the HTT phase, demonstrating an extremely high sample quality.  The high pressure x-ray diffraction experiment was performed on the triple-axis diffractometer at wiggler beamline BW5 at HASYLAB, using a piston-type pressure cell \cite{zimm08}.  The calibration of the pressure cells and estimation of pressure uncertainties are described in \cite{zimm08}.  To increase the signal to background ratio, a sample with optimized shape, 1.6~mm {\o} $\times$ 1.3~mm, was used. Taking advantage of the large penetration depth of 100-keV photons ($\lambda = 0.124$~\AA), the bulk properties of the charge stripe order as well as the crystal structure were studied in transmission geometry, using a $1\times 1$~mm$^2$ beam size.  Scattering vectors ${\bf Q}=(h,k,\ell)$ are specified in units of $(2\pi/a, 2\pi/a,2\pi/c)$, where $a=3.78$~$\rm \AA $ and $c=13.2$~$\rm \AA $  are the lattice parameters of the HTT unit cell.  Absolute intensities are normalized to a storage ring current of 100~mA. Charge stripe peak intensities, $I_{\rm CO}$, measured in several experimental runs, are normalized with $I \rm _{(206)}$(0~GPa)/$I_{(206)}(p)$ of the nearby (206) Bragg reflection.  The pressure dependence of the superconducting transition temperature $T_c$ was extracted from measurements of the AC-\sus , performed in a He-gas cell at pressures up to $p=0.58$~GPa, and in a diamond anvil cell with He pressure medium up to 14.7~GPa.

\begin{figure}[b]
\center{\includegraphics[width=0.95\columnwidth,angle=0,clip]{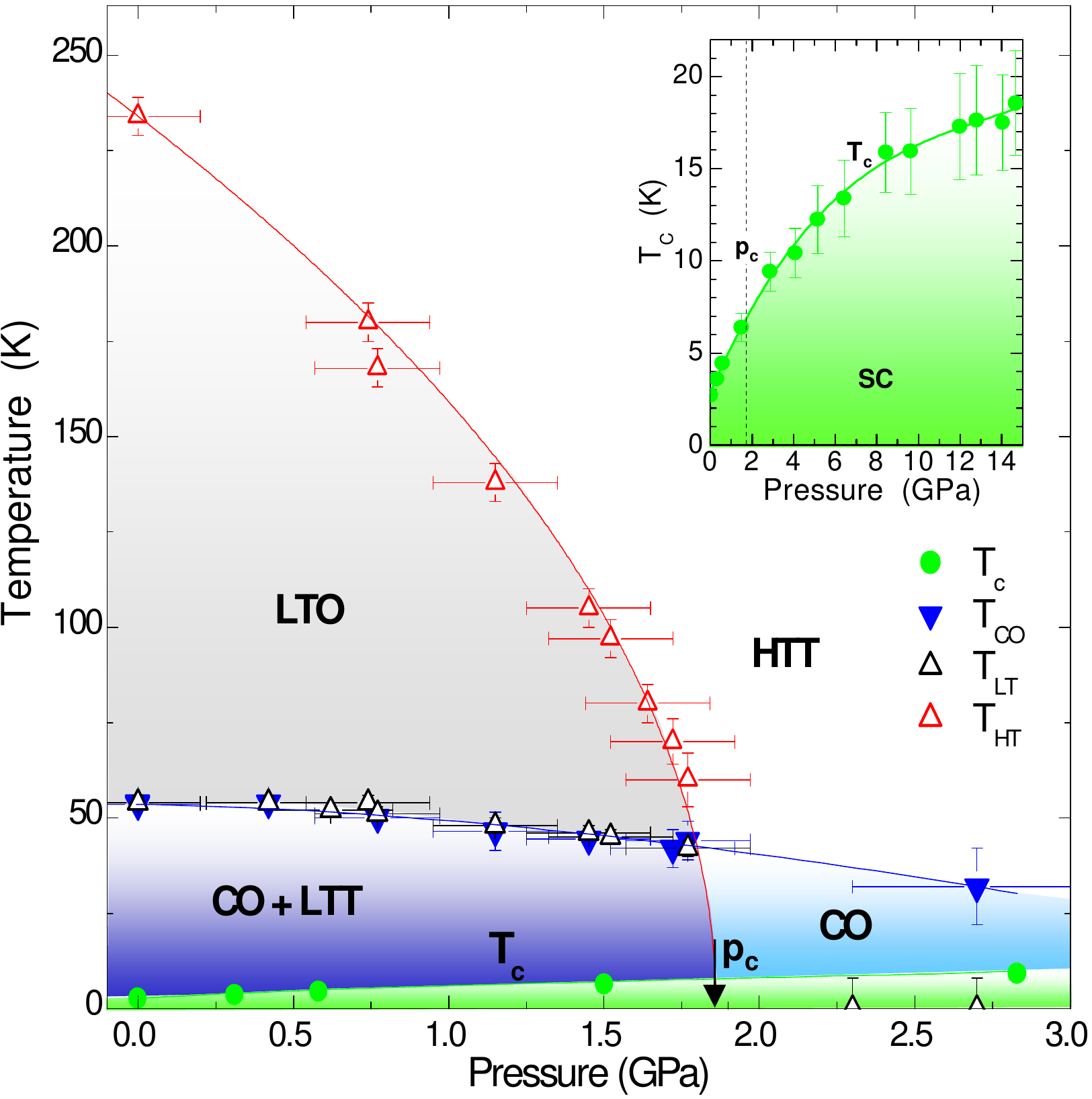}}
\caption{(color online).  Observed temperature vs.\ pressure phase diagram of
La$_{1.875}$Ba$_{0.125}$CuO$_4$. Indicated are the transition
temperatures of the structural phases HTT, LTO, and LTT, the charge stripe
phase (CO), and bulk superconductivity (SC). At ambient pressure we find $T_{\rm HT}=235$~K, $ T_{\rm LT}=54$~K, $T_{\rm CO}=54$~K and $ T_{c}=3$~K, respectively. The red line describes
$T_{\rm HT}(p)$ using $T_{\rm HT}(0)\cdot [(p_c-p)/p_c]^{0.5}$ with a critical pressure
of $p_c=1.85$. Inset: $T_c$ for pressures up to 14.7~GPa.} \label{fig1}
\end{figure}

The temperature versus pressure phase diagram in Fig.~\ref{fig1} summarizes our results.
Both the LTO and LTT phase are suppressed at $p_c=1.85$~GPa. Below $p_c$ the transition
temperature of the CO phase, $T_{\rm CO}$, and $T_{\rm LT}$ are locked together. Unlike the LTT phase, however, the CO phase continues to exist beyond $p_c$, with $T_{\rm CO}$
remaining higher than $T_c$. Within the 2.7~GPa
range of the diffraction experiment, $T_c$ increases from 3~K to 10~K. The inset of
Fig.~\ref{fig1} shows in more detail that, even at 14.7~GPa, $T_c$ reaches only 18~K
which is far below the maximum $T_c$ of $\sim 30$~K found in \lbco\ 
 \cite{mood88,yama92}.

\begin{figure}[t]
   \center{\includegraphics[width=1\columnwidth,angle=0,clip]{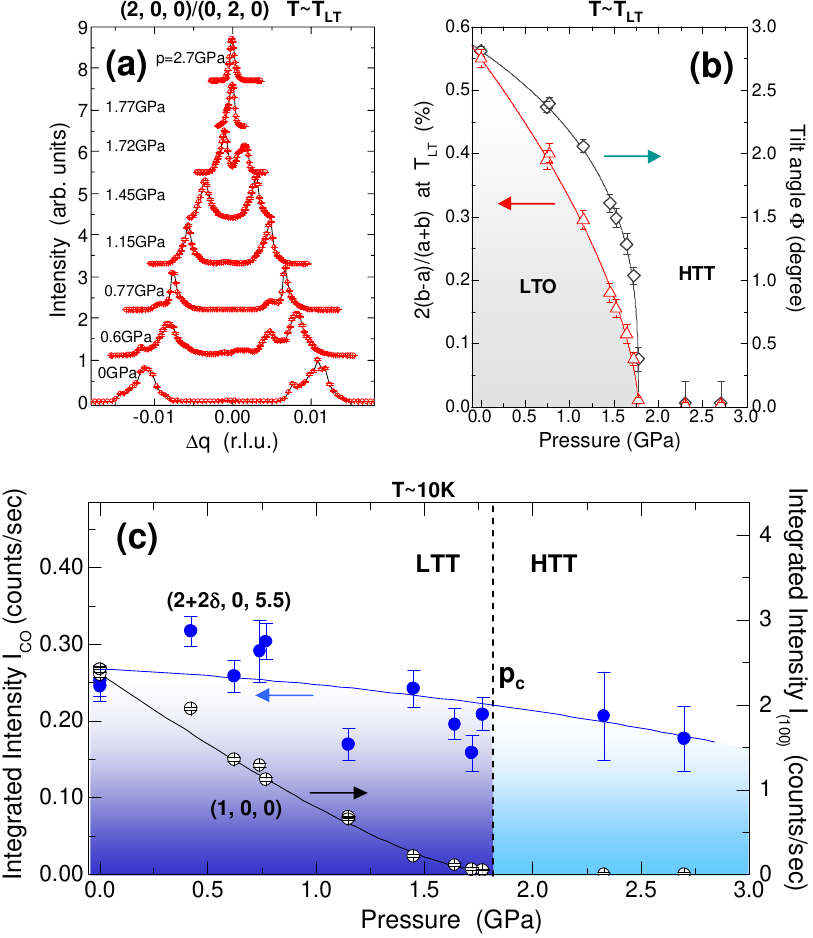}}
   \caption{(color online).  Pressure dependence of  crystal
   structure and charge stripe order.
    (a) Transverse scans at $T_{\rm LT}$ through the (200)
   and (020) Bragg reflections, showing the suppression
   of the orthorhombic splitting with pressure. (b) Orthorhombic strain 
   calculated from the splitting in (a), and average 
   tilt angle $\Phi$, calculated using
   $\Phi^2 = f \cdot (b-a)$ with $f=380$~$(^\circ)^2/{\rm \AA}$ \cite{craw05}.
   (c) Comparison of the integrated intensity from $k$-scans through the
   (100) LTT-peak and the $(2+2\delta,0,5.5)$ CO-peak. The vertical
   line indicates the critical pressure $p_c$.
 Solid lines are guides to the eye.}\label{fig2}
\end{figure}

To identify the different phases in Fig.~\ref{fig1}, we have performed scans in reciprocal
space through specific reflections; Fig.~\ref{fig2} presents key results.  Figure~\ref{fig2}(a) shows that the orthorhombic splitting between the (200) and (020) Bragg reflections (simultaneously present due to twin domains) is clearly resolved.  The pressure dependence of the orthorhombic strain, $2(b-a)/(a+b)$, at $T\agt T_{\rm LT}$ is shown in Fig.~\ref{fig2}(b), together with the calculated average tilt angle $\Phi$ of the CuO$_6$ octahedra.  

\begin{figure}[t]
\center{\includegraphics[width=1\columnwidth,angle=0,clip]{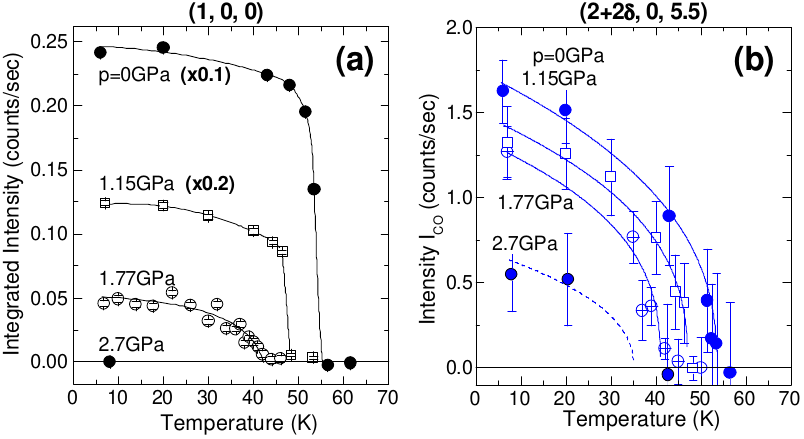}}
\caption{(color online).  Temperature and pressure dependence of LTT phase
and charge stripe order. (a) Integrated intensity from $k$-scans
through the (100) LTT peak. (b) Integrated intensity from $h$-scans through the
$(2+2\delta,0,5.5)$ CO-peak.  For the results at 2.7 GPa, in particular, the intensity vs.\ temperature is confirmed by $k$-scans.  Lines are guides to the eye.}\label{fig3}
\end{figure}

That the suppression of the average octahedral tilts at $p=p_c$ also occurs at our base temperature (10~K) is demonstrated by the decay of the (100) superlattice intensity in Fig.~\ref{fig2}(c); the (100) is unique to the LTT phase.  In sharp contrast, the figure shows that the intensity of the CO superlattice peak $(2+2\delta,0,5.5)$ decreases only modestly with pressure, remaining substantial at $p\gg p_c$ and indicating that CO survives in the phase with restored 4-fold symmetry.  (Note that the survival of the CO in the HTT phase has been observed at two pressures, 2.3 and 2.7 GPa, although the temperature-dependence was measured only at the second of these.)   The temperature-dependent data in Fig.~\ref{fig3}(b) demonstrate that $T_{\rm CO}$ also decreases only gradually with pressure.  The $T_{\rm LT}$ transition decreases at the same rate (for $p<p_c$), as indicated in Fig.~\ref{fig3}(a), but the amplitude of the LTT lattice modulation is strongly suppressed as $p\rightarrow p_c$.

The observation that $T_{\rm CO}=T_{\rm LT}$ for $p<p_c$ is consistent with a unidirectional charge modulation that couples to the in-plane anisotropy of the LTT phase.  In fact, without the presence of the stripe order, one might expect $T_{\rm LT}$ to decrease with pressure in proportion to $T_{\rm HT}$, as $\Phi\rightarrow0$.  With the detection of charge-stripe order in the HTT phase at $p>p_c$, one might expect that some local tilt distortions could be induced.  (Previous work has established that static octahedral tilt disorder is common in the HTT phase of \lbco\ \cite{bozi97,hask00,waki06}.) We show next that this is, in fact, the case.

\begin{figure}[t]
\center{\includegraphics[width=0.65\columnwidth,angle=0,clip]{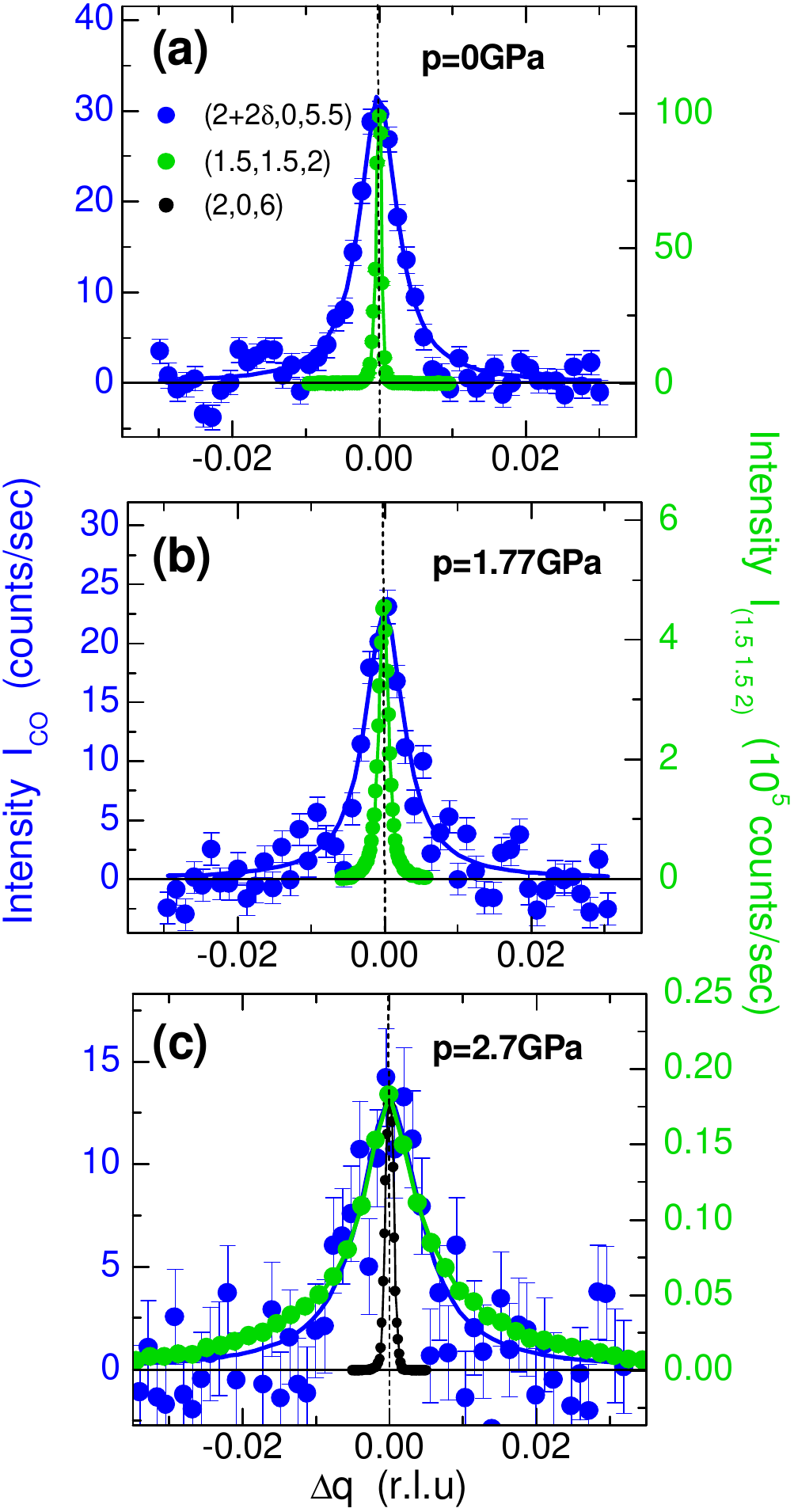}}
\caption{(color online).  Comparison of CO-peak and tilt-peak profiles.
$k$-scans through the $(2+2\delta,0,5.5)$ CO-peak and transverse $(h,k)$-scans through
the $(1.5,1.5,2)$ tilt-peak at $T\sim10$~K for representative pressures:
(a) $p=0$. (b) $p \alt p_c$. (c)
$p>p_c$. A linear background has been subtracted. Solid lines through the CO-peaks
are fitted Lorentzians. In (c) the $k$-scan through the (206) Bragg
reflection demonstrates the insignificance of pressure induced broadening at $p=2.7$~GPa.
The data in (c) were collected with a pressure cell with thicker walls
(stronger absorption), resulting in lower counting statistics.}\label{fig5}
\end{figure}

Figure~\ref{fig5} shows transverse scans of the $(2+2\delta,0,5.5)$ CO peak at pressures from below to above $p_c$.  The CO peak has a finite width at ambient pressure \cite{tran08}, with the inferred correlation length $\xi$ decreasing only slightly from 120~\AA\ at 0~GPa to 80~\AA\ at 2.7~GPa.  Ideally, we would like to test at $p>p_c$ for possible diffuse scattering at a superlattice position such as (100) that is unique to the LTT phase; unfortunately, those features have very small structure factors.  Instead, we have followed the $(1.5,1.5,2)$ peak, which is sensitive to octahedral tilts in both the LTO and LTT phases.  Transverse scans through that position are indicated in green in Fig.~\ref{fig5}.  At ambient pressure, the peak is resolution limited, while it has developed a small but finite width at 1.77~GPa ($\xi\sim500$~\AA).  Finally, at 2.7~GPa, where there is no long-range order associated with octahedral tilts, the width of the scattering centered at $(1.5,1.5,2)$ matches that of the CO peak.  The width of the weak, diffuse $(1.5,1.5,2)$ peak is found to saturate at its minimum value for $T<T_{\rm CO}$.   

\begin{figure}[t]
\center{\includegraphics[width=0.8\columnwidth,angle=0,clip]{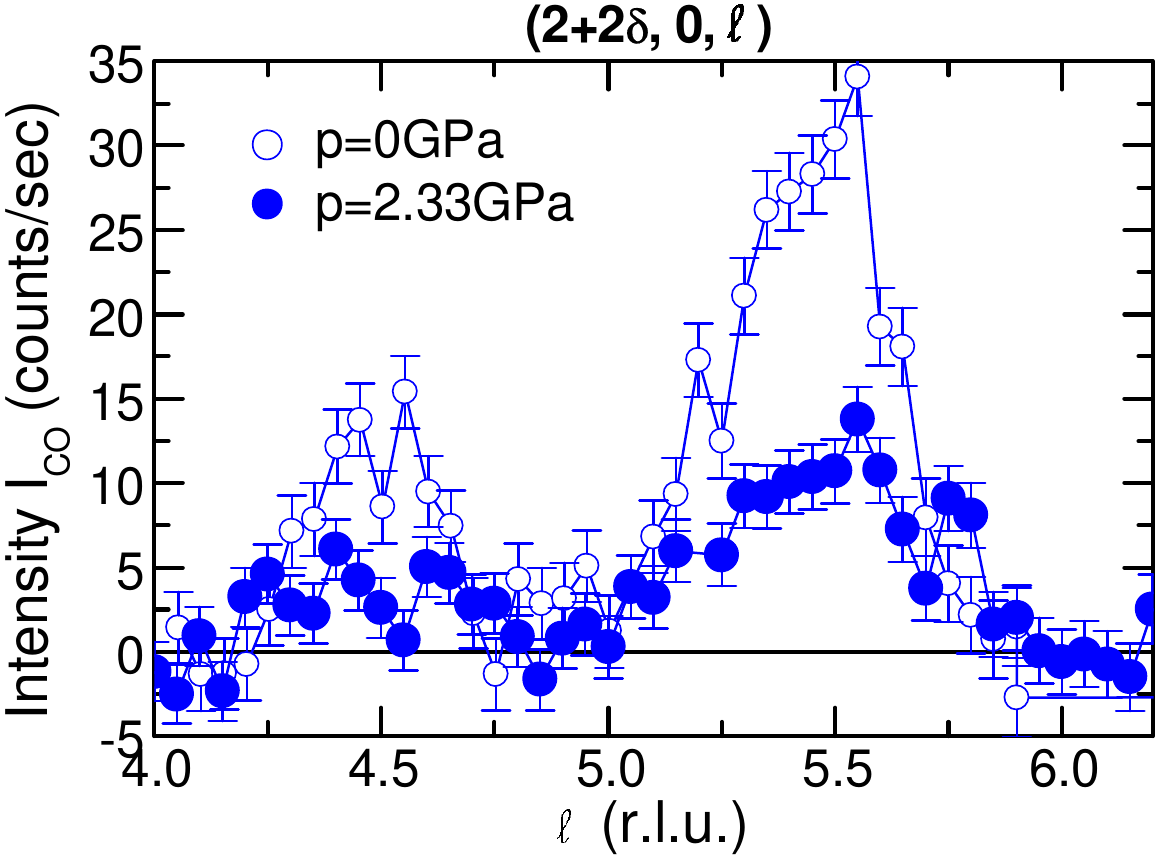}}
\caption{(color online).  Background-subtracted intensity of the CO intensity measured along ${\bf Q}=(2+2\delta,0,\ell)$ at base temperature for $p=0$~GPa (open circles) and $p=2.33$~GPa (filled circles). In the first case, background was measured along the same {\bf Q} at $T=60$~K; in the second, background was measured at base temperature along a parallel path in {\bf Q} displaced by $(0,0.03,0)$, corresponding to a displacement of several peak widths in $k$.}\label{fig4}
\end{figure}

Based on the scattering data, we come to the conclusion that charge stripes in the HTT phase are consistent with a short range ($\xi\sim5$ stripe periods)
smectic electronic liquid-crystal state \cite{kive98} that breaks the
rotational symmetry and freezes in the high pressure regime due to its coupling to local
octahedral tilts.  The smooth variation of the stripe order through $p_c$ is also consistent with the gradual rise in superconducting $T_c$, as indicated in Fig.~\ref{fig1}.  The depression of the bulk $T_c$ by stripe order at ambient pressure is known to be due to a frustration of interlayer coupling, as their is evidence of two-dimensional superconducting correlations at temperatures as high as 40~K \cite{li07,tran08}.  It has been proposed that the frustration could be due to a type of striped superconducting state \cite{hime02,berg07}, also described as a pair-density-wave state \cite{berg09b}.  That proposal depends on the 90$^\circ$ rotation of the stripes from one layer to the next that is associated with stripe pinning in the LTT phase \cite{vonz98}.  Figure~\ref{fig4} shows scans of the CO intensity over a range of {\bf Q} varying perpendicular to the planes.  At high pressure we observe the same sinusoidal modulation (with zeroes at integer $\ell$) as at ambient, demonstrating that in the HTT phase the charge stripes retain the interlayer correlations found in the LTT.

It was noted quite some time ago that the impact of pressure on $T_c$ in \lbco\ is very sensitive to $x$ \cite{yama92b}.  In particular, $T_c$ rises very slowly with $p$ for $x=\frac18$, but quite rapidly for $x$ slightly larger or smaller than $\frac18$.  Compared to our results, Crawford {\it et al.} \cite{craw05} found a more rapid rise in $T_c$ near $p_c$ in La$_{1.48}$Nd$_{0.4}$Sr$_{0.12}$CuO$_4$, which might be due to the small nominal difference in hole doping.  Takeshita {\it et al.} observed an extreme sensitivity of $T_c$ to strain along [110] in La$_{1.64}$Eu$_{0.2}$Sr$_{0.16}$CuO$_4$, but this could be due to inducing a transition from LTT to LTO.   The survival of charge-stripe order in the HTT phase of \lbcoe, along with extreme sensitivity of $T_c(p)$ to $x$, may indicate a special stability at $x=\frac18$ for the order (such as the proposed PDW state) that decouples the layers.

We gratefully acknowledge helpful discussions with E. W. Carlson, S. A. Kivelson, and A. Gozar.
M.v.Z. and M.H. thank R. Nowak for technical support with the
high pressure XRD setup. Work at Brookhaven is supported by the Office of Science, US
Department of Energy under Contract No.\ DE-AC02-98CH10886. M.D. and J.S.S. are supported by the National Science Foundation through Grant No. DMR-0703896.


\end{document}